\documentclass[final,1p,times]{elsarticle} 
\usepackage{graphicx} 
\usepackage{amssymb} 
\usepackage{amsthm} 

\journal{Nuclear Physics A} 
\begin{document} 

\begin{frontmatter} 


\title{High $p_T$ Parton Energy Loss from Jet Pair Correlation Observables in the PHENIX Experiment}

\author{Chin-Hao Chen$^{a}$ for the PHENIX collaboration}

\address[a]{Department of Physics and Astronomy, Stony Brook University, 
Stony Brook, NY 11794-3800, USA}

\begin{abstract} 
By controlling several experimental observables, such as momentum and reaction plane, the PHENIX experiment has performed a systematic study of high $p_T$ parton energy loss in hot nuclear matter, as well as jet-induced medium modifications, such as the ridge and shoulder. Baseline d+Au measurements are also studied to provide the comparison as a control for cold nuclear matter effect. 
\end{abstract} 

\end{frontmatter} 




One of the most striking results made at RHIC is jet quenching in the hot dense nuclear matter. By using two particle correlations, we are able to study how high $p_T$ partons lose energy in the medium. When triggering on high $p_T$ particles, the opposing or away-side jet shape varies with associated $p_T$. At high associated $p_T$, an away-side punch through jet is observed, which means that some part of the jet survived the medium. At intermediate and low associated $p_T$, the away-side displays a double peak structure, instead of a single peak at $\Delta\phi \approx \pi$. Measurements in p+p and d+Au are used as benchmarks: p+p measures hard scattering with no medium, while d+Au determines cold nuclear matter effects. An important question is where the lost energy goes. When we look at low $p_T$ correlations, we find that not only is the away-side modified, but the near-side is modified as well. The near-side has an enhancement along $\Delta\eta$, or the ``ridge''. The away-side has a double peak structure, which is the ``shoulder'' ~\cite{PPG083}. 



In d+Au collisions, we use non-identified charged hadrons as both the trigger and associated particles and study the two particle azimuthal correlations ~\cite{Jia}. The result is compared with p+p collisions. The angular widths of jets in d+Au collisions are consistent with these in p+p collisions. Comparison of the yields in p+p and d+Au is characterized by the ratio of the per trigger yield (PTY) in d+Au to the per trigger yield in p+p, or $I_{dAu} = (PTY_{dAu})/(PTY_{pp})$. At high trigger $p_T$ ($>$ 2 GeV/c), $I_{dAu}$ approaches 1 indicates that when both trigger and associated particles are from hard scattering, the yields are consistent with p+p. In d+Au collisions, there is more enhancement in baryon than in meson ~\cite{PPG028}. So further correlation study with particle identification can help to quantify the initial state effect on the particle production mechanism.  

The observations in d+Au show that the modification of the jet is due to the medium. To quantify this, we use high $p_T$ $\pi^{0}$ triggers and compare central Au+Au collisions (centrality 0-20\%) with p+p results. There are several advantages to using $\pi^{0}$ as trigger particles. First, $\pi^{0}$ are well reconstructed in PHENIX with less background at high $p_T$ than charged hadrons. Second, it may be easier to interpret the results with identified than non identified hadron triggers because of the baryon anomaly at intermediate $p_T$ ~\cite{PPG015}. 



\begin{figure}[ht]
\centering
\includegraphics[width=1.0\textwidth]{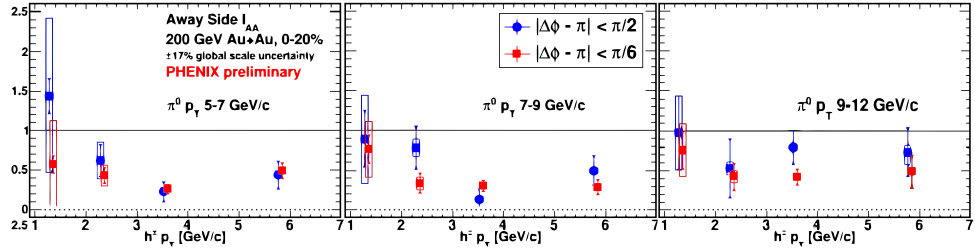}	 
\caption[]{Away-side $I_{AA}$ as a function of partner $p_T$ in three trigger $p_T$ bins in $\pi^{0}-h^{\pm}$.}
\label{fig:pi0IAA}
\end{figure}

We study $\pi^{0}$-h correlations in central Au+Au and p+p by examing the yields in the away-side ~\cite{Adare}. We use $I_{AA} = (PTY_{AA})/(PTY_{pp})$ to describe the away-side yield. Fig.~{\ref{fig:pi0IAA}} shows $I_{AA}$ vs partner $p_T$ in three different trigger $p_T$ bins. The away-side $I_{AA}$ are measured in two different angular ranges, $|\Delta\phi-\pi|<\pi/2$ and $|\Delta\phi-\pi|<\pi/6$, corresponding to the entire away-side and the punch-through jet region respectively. We observe a similar suppression pattern at associated particle $p_T$ $>$ 2 GeV/c in both the head and total away-side region. Also, the away-side width in Au+Au is broader than in p+p at $\pi^{0}$  $p_T$ = 5-7 GeV/c but not as broad at higher $\pi^{0}$ $p_T$. ~\cite{Adare}. So the away-side shoulder structure is not obvious within the large error bar in present data at high $\pi^{0}$ $p_T$. This may indicate that the whole away-side is suppressed as much as the punch through jet with high $p_T$ triggers. 

In order to further explore the away-side suppression shown above, we vary the angle of the $\pi^{0}$ with respect to the reaction plane. In mid-central collisions, we are then able to vary the average path length of the away-side jet through the matter. We define the angle between the trigger particle and the reaction plane as $\phi_{S}$, with $\phi_{S} = 0 $ in-plane, and $\phi_{S} =\pi$ out-of-plane. Two particle correlations as a function of $\phi_{S}$ provides a sensitive probe of the medium length dependence of the parton medium interaction. 

\begin{figure}[ht]
\begin{minipage}[b]{0.45\linewidth}
\centering
\includegraphics[width=0.90\textwidth]{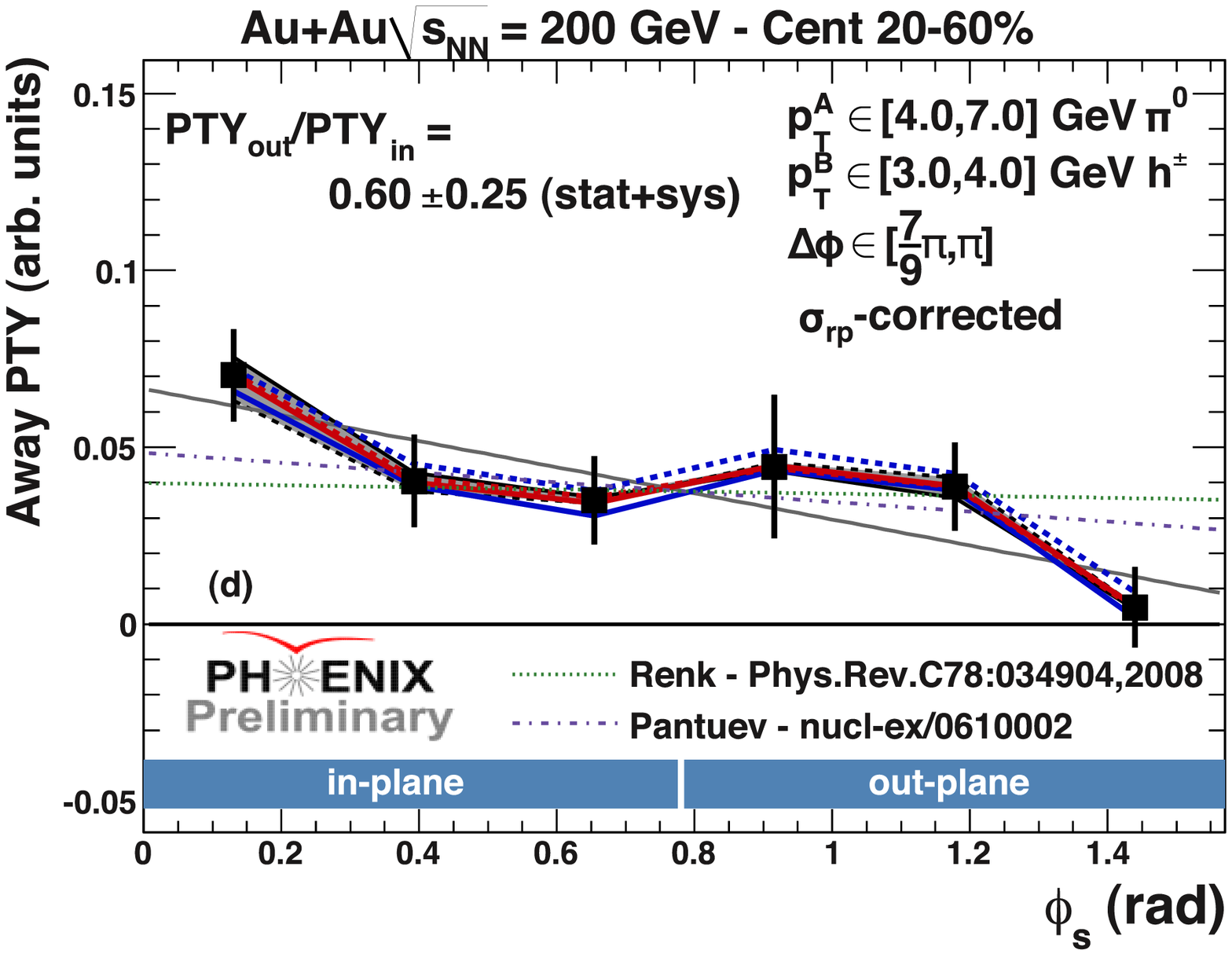}	       
\caption[]{Away-side per trigger yield vs $\phi_{S}$ in $\pi^{0}-h^{\pm}$}
\label{fig:yield_vs_phis}\end{minipage}
\hspace{0.2cm}
\begin{minipage}[b]{0.45\linewidth}
\centering
\includegraphics[width=0.90\textwidth]{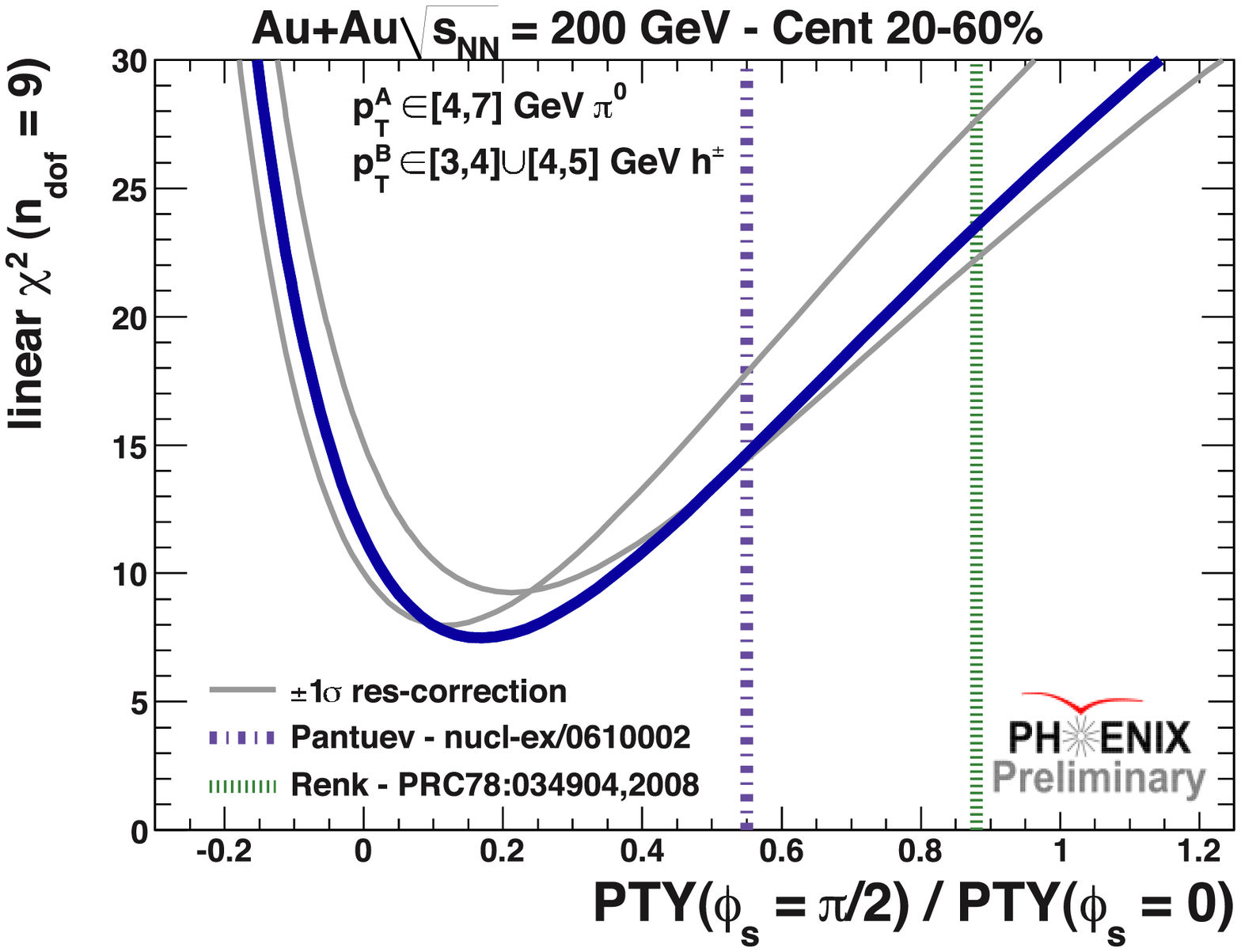}	 
\caption[]{$\chi^{2}$ value vs in-plane to out-of-plane per trigger yield ratio.}
\label{fig:chi2_plot}
\end{minipage}
\end{figure}

We use 4-7 GeV/c $\pi^{0}$ triggers, and 3-4 GeV/c non-identified charged hadron as associated particles to measure the correlations as a function of $\phi_{s}$ ~\cite{McCumber}. In the 20-60\% centrality bin, the path lengh variation is stronger than central collisions from in to out-of-plane. Fig.~{\ref{fig:yield_vs_phis}} shows the away-side per trigger yields (PTY) are measured as a function of $\phi_{s}$. The away-side yield decreases with increasing $\phi_{s}$, indicating that the away- side yield decreases from in to out-of-plane. A similar trend is observed at higher associated $p_T$ (4-5 GeV/c). 

We fit the $\phi_{S}$ dependence with a line, and plot the $\chi^{2}$ of the fit as a function of the ratio of ($PTY|_{\phi_{S}=\pi}$)/($PTY|_{\phi_{S}=0}$). The results are in Fig.~{\ref{fig:chi2_plot}}. The minimum $\chi^{2}$ is at ($PTY|_{\phi_{S}=\pi}$)/($PTY|_{\phi_{S}=0}$) $\approx$ 0.2, which means the out-of-plane yield is suppressed by a factor of 5 from the in-plane yield. Two theory calculations ~\cite{Pantuev, Renk} predict the away-side suppression, but these calculations have less suppression in out-of-plane than the data indicates. 

\begin{figure}[ht]
\begin{minipage}[b]{0.49\linewidth}
\centering
\includegraphics[width=0.95\textwidth]{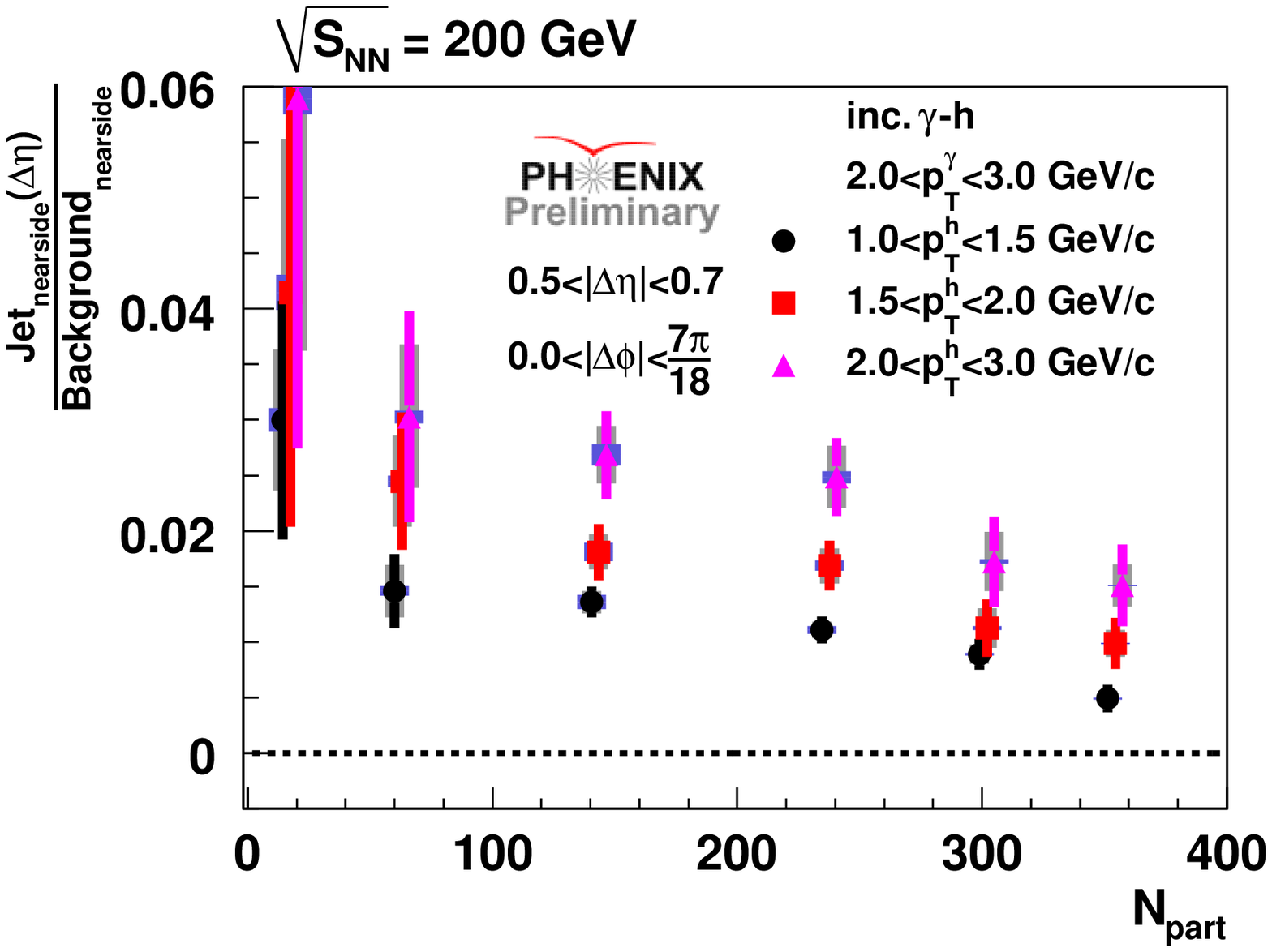}
\caption{Ridge to background ratio vs $N_{part}$}
\label{fig:nr_ratio_flow_deta_50_70}
\end{minipage}
\begin{minipage}[b]{0.49\linewidth}
\centering
\includegraphics[width=0.95\textwidth]{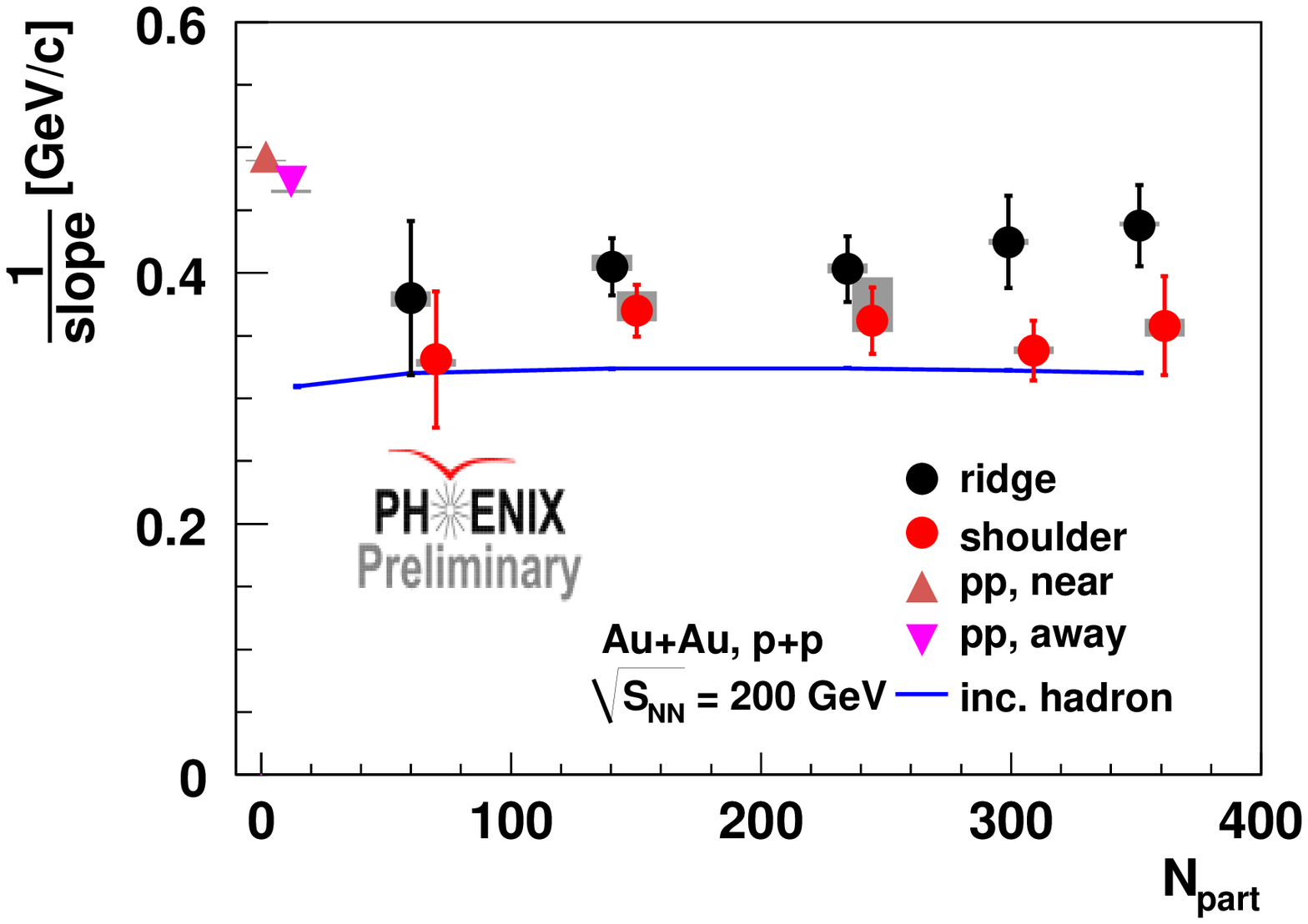}
\caption{Inverse slope of ridge and shoulder vs $N_{part}$.}
\label{fig:slope_result}
\end{minipage}
\end{figure}

When both trigger and associate particle $p_T$ are lowered, we begin probing the medium response to the jets. In heavy ion collisions, there is an enhancement, known as the ridge, along $\Delta\eta$ at $\Delta\phi$ $\approx 0$ in two particle $\Delta\eta$-$\Delta\phi$ correlations. We use inclusive photons at 2-3 GeV/c as trigger particles and correlate with non-identified charged particles. At this photon $p_T$ range, the inclusive photons are mostly from meson decay. In heavy ion collisions, the multiplicity increases with the number of participants, $N_{part}$. If the ridge is directly related only to the combinatoric background, then the yield of ridge should scale with the underlying events. Fig.~{\ref{fig:nr_ratio_flow_deta_50_70}} shows the ratio between the ridge yield and the yield of the underlying event, as a function of partner $p_T$ and $N_{part}$. The ridge yield is a few percent of the underlying events, but for $N_{part} > 200$, the ratio decreases with increasing $N_{part}$. This means that the ridge yield does not increase as fast as the number of particles in the underlying event. The ratio increases with $p_T$ indicating that the ridge is harder than the underlying event. Consequently, the ridge can not be due to combinatorial background.  


The away-side has a double peak structure in central Au+Au collisions. We use three Gaussians to fit the away-side and separate the different components. The punch-through jet is described by a Gaussian peak at $\Delta\phi = \pi$. Two additional Gaussian fits, symmetric about $\pi$, are used to describe the jet modification, known as the shoulders. The shoulder yield is the sum of the two Gaussian fits. The spectra of ridge and shoulder in different centralities are used to compare the medium modification in both near and away-side. Fig.~{\ref{fig:slope_result}} shows the inverse slope extracted from the spectra of the ridge and shoulder as a function of $N_{part}$. This similarity may indicate that both the ridge and shoulder come from a similar mechanism. The inverse slopes of ridge and shoulder are compared with hard scattering, or p+p, and inclusive charged hadron, from the medium. Both are softer than hard scattering and are slightly harder than the medium. Consequently, neither the ridge nor the shoulder come directly from hard processes, because they are much softer than p+p. This will give us information on how semi-hard particles interact with the matter. If both ridge and shoulder are from the medium, there must be some mechanism to excite them, to cause the spectra to be harder than that of the medium.

\begin{figure}[ht]
\centering
\includegraphics[width=0.49\textwidth]{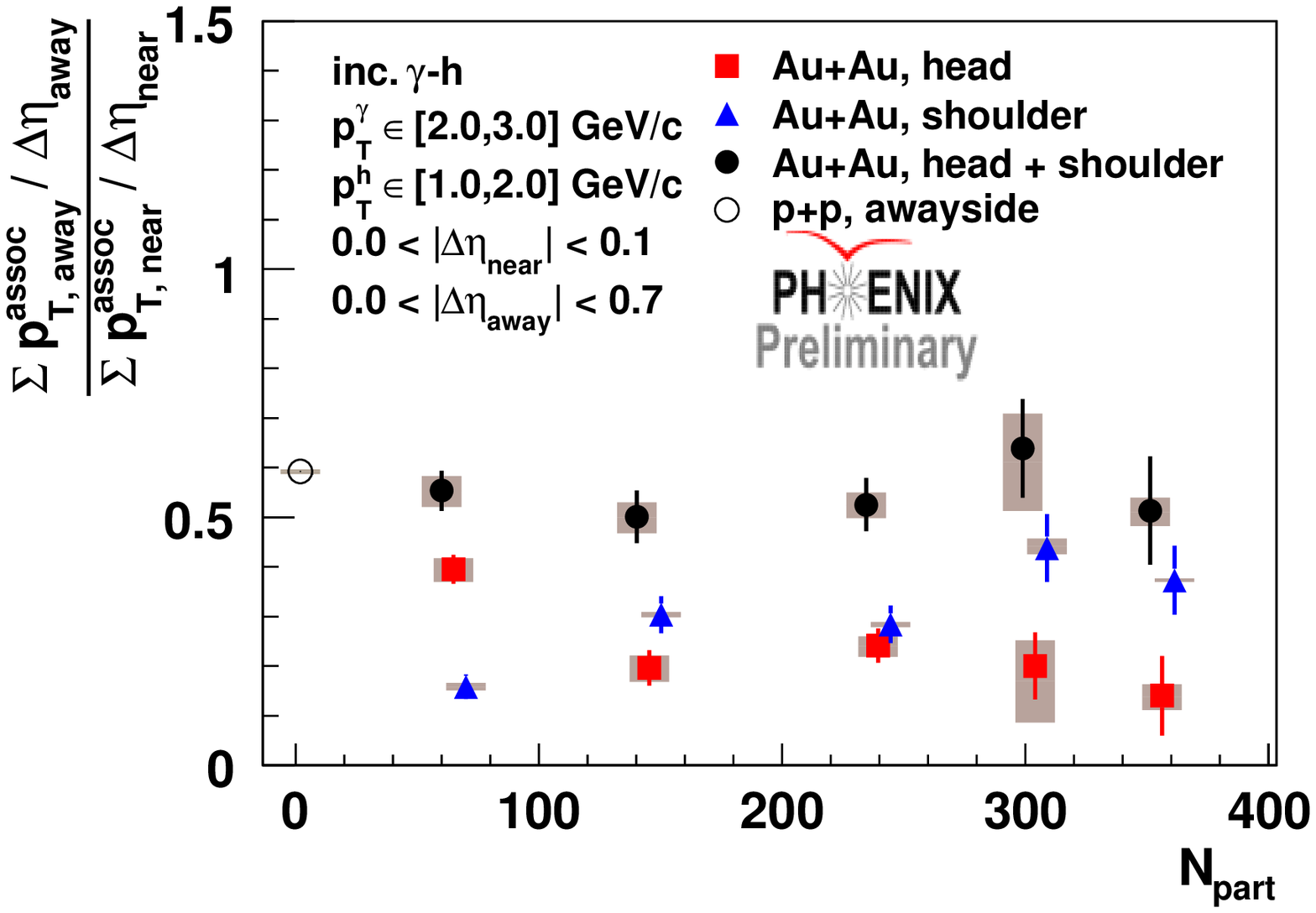}	 
\includegraphics[width=0.49\textwidth]{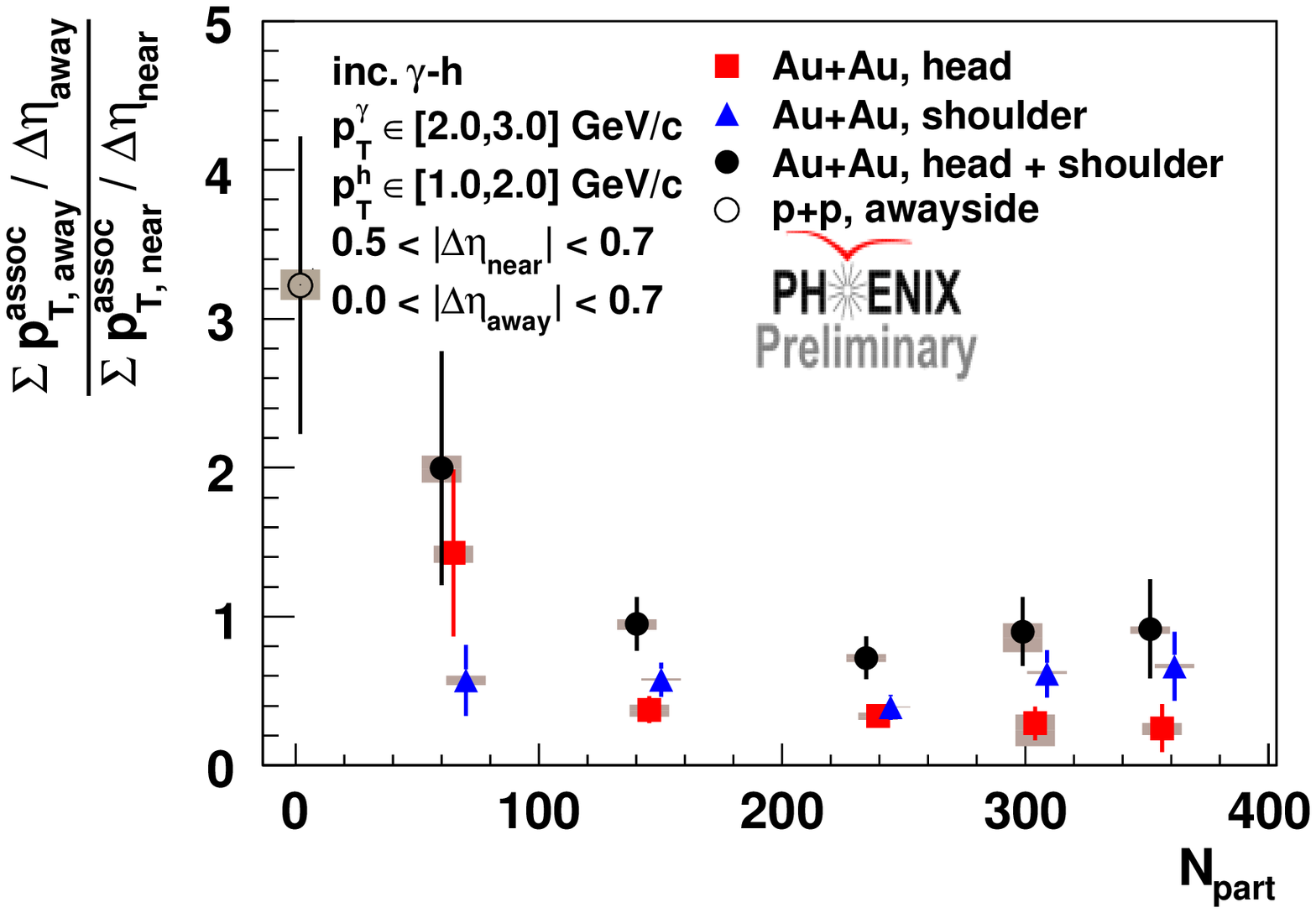}	       
\caption[]{Ratio of away-side $p_T$ over near-side $p_T$ vs $N_{part}$ at $0.0<|\Delta\eta|<0.1$ and $0.5<|\Delta\eta|<0.7$}
\label{fig:pT_ratio}
\end{figure}

If the away-side jet is quenched in the medium, conservation of transverse momentum requires that the momentum must come out somewhere. The near and away-side yield is weighted with transverse momentum along the direction of the trigger particle to investigate how the $p_T$ is carried by the associated particles in the near and away-sides. Here we did not include the $p_T$ of the trigger particle. We take the ratio between away and near-side weighted yields as a function of $N_{part}$ in different $\Delta\eta$ regions and plot the result in Fig.~{\ref{fig:pT_ratio}}. At $0.0<|\Delta\eta|<0.1$, the near-side is mostly jet. Both near and away-side yields increase with $N_{part}$, but the away to near-side ratio of total $p_T$ is approximately constant with $N_{part}$ This value of 0.55 is consistent with p+p, indicating that the $p_T$ observed in the near and away-side jets increase together with collision centrality. As the collisions become more central, the fraction of the transverse momentum carried by the head region of the away-side jet decreases, while that in the shoulder increases. This seems to suggest that transverse momentum lost from the scattered parton is transferred to the shoulder region.


At $0.5<|\Delta\eta|<0.7$, the ratios between the away and near-sides are large in p+p and peripheral collisions because of very little near-side yield. At more central collisions, the near-side increases because the ridge is the dominant source of particles in this $\Delta\eta$ range. The associated particle $p_T$ ratio between shoulder and ridge hold constant along $N_{part}$ at about 0.5. This says that the ridge carries about twice the $p_T$ of the shoulder. 

In summary, in central Au+Au collision, at high $p_T$, the high $p_T$ parton loses energy in the medium and has a strong reaction plane dependence. At low $p_T$, the ridge and shoulder have similar inverse slope, which may indicate both have the same origin. The ratio of the $p_T$ carried by the near-side and away-side associated particles at $\Delta\eta \approx 0$ indicates that the $p_T$ loss in the medium comes out from the shoulder. The new high-statistics data sets will allow us to perform more detailed analysis, such as high $p_T$ particle correlation with particle identification, to study the parton energy loss in the medium. 




\end{document}